\documentclass[iop]{emulateapj}
\usepackage{graphicx}
\usepackage{epstopdf}
\usepackage{natbib}
\usepackage{xspace}
\usepackage{color}

\newcommand{\iras}   	{IRAS~16293$-$2422\xspace}
\newcommand{\irasA}   	{I16293A\xspace}
\newcommand{\irasB}   	{I16293B\xspace}
\newcommand{\kms}		{km~s$^{-1}$\xspace}
\newcommand{\JyB}		{Jy~beam$^{-1}$\xspace}
\newcommand{\msun}		{$M_{\sun}$\xspace}
\newcommand{\lsun}		{$L_{\sun}$\xspace}
\newcommand{\Pdot}		{\msun~\kms~yr$^{-1}$}
\newcommand{\Mdot}		{\msun~yr$^{-1}$}

\newcommand{\vlsr}		{$v_{\rm LSR}$\xspace}
\newcommand{\ie}    		{i.\,e.,}

\newcommand{\cts}		{C$^{34}$S}

\begin{document}

\title{On the origin of the molecular outflows in \iras}

\author{Josep M. Girart\altaffilmark{1}, Robert Estalella\altaffilmark{2}, 
Aina Palau\altaffilmark{1}, Jos\'e M. Torrelles\altaffilmark{1,2}, 
Ramprasad Rao \altaffilmark{3} 
}

\affil{$^1$  Institut de Ci\`encies de l'Espai, (CSIC-IEEC), Campus UAB,
Facultat de Ci\`encies, C5p 2, 08193 Bellaterra, Catalonia, Spain,
girart@ice.cat}
\affil{$^2$ Departament d$'$Astronomia i Meteorologia, Institut de Ci\`encies
del Cosmos (UB-IEEC), Mart\' i i Franqu\`es, 
Universitat de Barcelona, 08028 Barcelona, Catalonia, Spain}
\affil{$^3$ Institute of Astronomy and Astrophysics, Academia Sinica, 645
N. Aohoku Pl., Hilo, HI 96720, USA}

\begin{abstract}
We present CO 3--2, SiO 8--7, \cts\ 7--6, and 878~$\mu$m dust continuum 
subarcsecond angular resolution observations with the Submillimeter Array (SMA) 
toward the IRAS~16293$-$2422 (I16293) multiple low-mass protostellar system.  
The \cts\ emission traces the 878~$\mu$m dust continuum well, and in addition 
clearly shows a smooth velocity gradient along the major  axis of component \irasA.  
CO shows emission at moderate high velocities arising from two bipolar outflows, 
which appear to be perpendicular with respect to each other.  The high sensitivity 
and higher angular resolution of these observations allows us to pinpoint well the 
origin of these two outflows at the center of component \irasA. Interestingly, the 
most compact outflow appears to point toward \irasB.  Our data show that the 
previously reported  monopolar blueshifted CO outflow associated with component 
\irasB\ seems to be part of the compact outflow arising from component \irasA. 
In addition, the SiO emission is also tracing this compact outflow: on one hand, 
the SiO emission appears to have a jet-like morphology along the southern 
redshifted lobe; on the other hand, the SiO emission associated with the 
blueshifted northern lobe traces a well defined arc on the border of component \irasB\ 
facing \irasA. The blueshifted  CO lobe of the compact outflow splits into two lobes 
around the position of this  SiO arc. All these results lead us to propose that the 
compact outflow from component \irasA\  is impacting on the circumstellar gas 
around component \irasB, possibly being diverged  as a consequence of the interaction. 
\end{abstract}


\section{Introduction}\label{intro}

The dark cloud Lynds 1689N, located in the Ophiucus star-forming region at a
distance of 120 pc \citep{Knude98, Loinard08, Lombardi08}, harbors \iras, one of
the best studied low-mass (class 0) protostellar systems \citep[hereafter
I16293; see][and references therein]{Alves12, Kristensen13, Loinard13, Zapata13}.  This 
system has two main components, \irasA\ and \irasB, separated by $\sim$ 5$''$ ($\sim$
600 AU), first detected at cm-continuum wavelengths \citep{Wootten89, Estalella91} 
and later also detected at (sub)mm wavelengths \citep[e.g.,][]{Chandler05, Rodriguez05,
Rao09, Pineda12, Loinard13}.  While \irasA\ shows an ``hourglass'' magnetic field 
structure, \irasB\ shows an ordered magnetic field \citep{Rao09}.  
\citet{Wootten89} found that \irasA\ splits into two subcomponents at 
cm wavelengths (usually referred as A1 and A2), separated by $0\farcs3$ 
(36~AU) and tracing two stars in a binary system \citep[e.g., ][]{Loinard07}.  
H$_2$O maser emission is also observed toward this binary \citep{Wilking87, 
Wootten89, Imai07}, tracing zones of compressed gas produced by shocks in the 
presence of very strong line-of-sight magnetic fields \citep[$\sim$ 113 mG, ][]{Alves12}. 
I16293 has two bipolar outflows at scales of $\sim 0.1$~pc 
\citep{Walker88, Mizuno90}, one of them centered on \irasA 
\citep{Yeh08}. A third, more compact bipolar molecular outflow has been also 
reported through Submillimeter Array (SMA) observations, centered 
on \irasA\ and extending along an axis through \irasA\ and \irasB\ 
\citep[SE-NW direction;][]{Rao09}. The observations indicate that 
most of the outflow activity in this region is concentrated on the sources  within \irasA,
with little outflow activity (if any) in the nearby \irasB\ component, where a
compact, possibly isolated protoplanetary disk around a protostar has been
inferred \citep{Rodriguez05}.  However, very recent ALMA CO 6--5 observations with 
an angular resolution of $\sim$ 0.3$''$ led \citet{Loinard13}
to suggest that \irasB\ is ejecting a blueshifted bubble-like outflow with 
low velocity and moderate collimation.  Based on these results, together with
the small kinematic age estimated for this outflow, these authors proposed that
\irasB\ is the youngest object in the region, and one of the youngest
protostars known  \citep{Loinard13}. 

In this Letter, we present new SMA CO 3--2, SiO 8--7, \cts\ 7--6, and continuum
observations at 345 GHz toward I16293. Our observations are more sensitive 
to extended structures ($\sim$ 20$''$) than those of the recently reported 690 GHz 
ALMA observations (the visibility range for ALMA is 62--943~k$\lambda$ and 
for the SMA is 20-240~k$\lambda$). 
Due to these observational properties, our data show that the blueshifted
outflow reported previously with ALMA data seems to be part of a bipolar outflow
originating from one of the stars within \irasA, rather than 
originating from \irasB\ as was recently proposed. More importantly, this bipolar 
outflow is interacting with \irasB\ producing a shock structure at its SW 
edge seen in SiO.

\section{Observations}

The SMA observations were taken on 2010 August 28 in the extended configuration.
These observations were performed in the polarimetric mode. The results of the polarization
data have been presented in a different paper \citep{Rao13}.
The receiver was tuned to cover the 333.5-337.5~GHz and  345.5-349.5~GHz
frequencies in the lower side band (LSB) and upper side band, (USB) respectively. The
phase center of the telescope was  RA(J2000.0)$=16^{\rm h}32^{\rm m}22\fs90$
and  DEC(J2000.0)$= -24\degr28\arcmin 35\farcs73$.  The correlator provided a
spectral resolution of about 0.8~MHz (i.e., 0.7~km~s$^{-1}$ at 345~GHz).  The
gain and bandpass calibrators were QSO J1733-130 and QSO 3C454.3, respectively. 
The absolute flux scale was determined from observations of Neptune. The 
flux uncertainty was estimated to be $\sim20$\%.
The data were reduced using the MIRIAD software package \citep{Wright93}. 
Self-calibration was performed independently for the USB and LSB on the
continuum emission of I16293.  We have selected the CO 3--2 (345.796 GHz) and 
SiO 8--7 lines (347.331 GHz), to trace the outflow activity in I16293, and the 
\cts\ 7--6  (337.396 GHz) line to trace the circumstellar gas in the region. 
In addition, the dust continuum at 878~$\mu$m is also presented.  Maps were 
obtained from the visibilities using natural weighting, which yielded a 
synthetic beam size of $\simeq 0\farcs8$,  and
allowed the tracing of smaller spatial scales ($\simeq100$~AU) around I16293 than 
the scales ($\simeq 300$~AU) traced by \citet{Rao09}.

\begin{figure}[h]
\epsscale{1.0}
\includegraphics[width=\columnwidth]{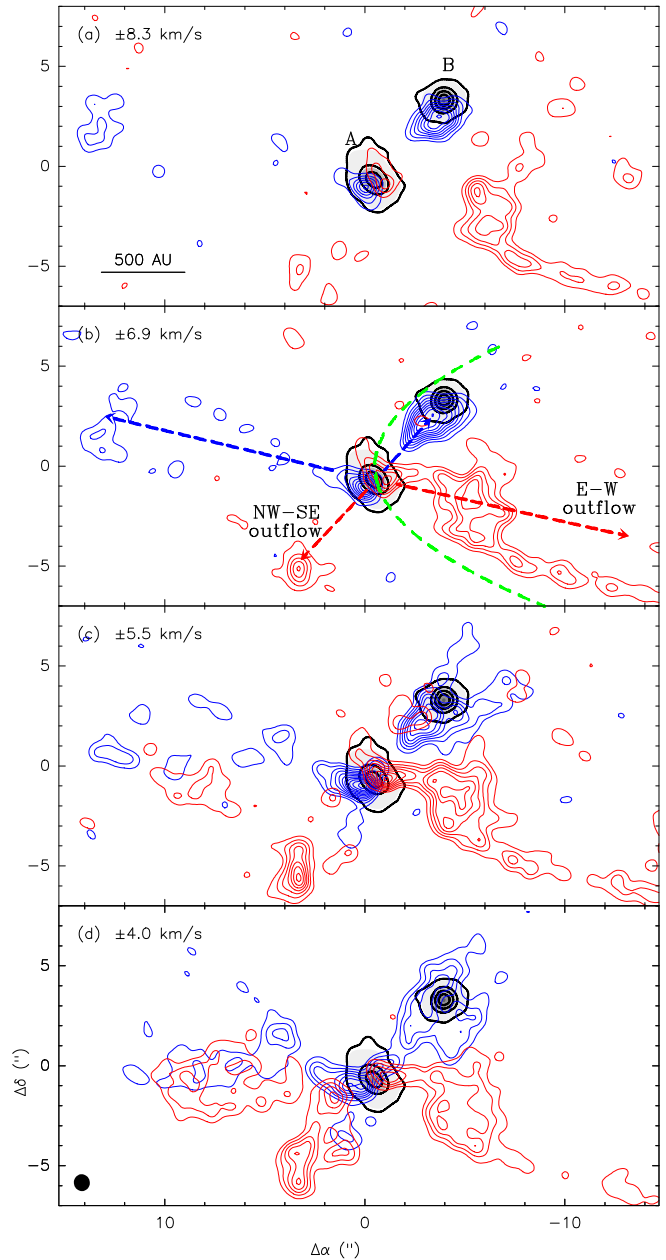}
\caption{Channel maps of the CO 3-2 for the blueshifted (blue contours) and
redshifted  (red contours) emission, overlapped with the 878~$\mu$m dust
emission (black  contours and grey scale). Panels $a$, $b$ and $c$: the CO 
contour's step level and the first contour are 0.47~\JyB. Panel $d$: 
the CO contour's step level as well the  first contour is 0.65~\JyB. 
In all panels the first contour is at 2-$\sigma$ level. The dust 
contours show the emission at the 5, 30, 55, and 80\%  level of the maximum intensity,
1.88\JyB.  The outflow velocity (\ie\ the velocity of the  gas with respect to the
systemic velocity of \irasA, $\simeq 3.5$\kms) of each 
channel is indicated in the top left corner of each panel. The offset spatial 
positions are with respect to the phase center (given in Section 2).
The green dashed line in panel $b$ shows the cavity traced by the E-W CO lobe at scales 
of $\sim 3000$~AU \citep{Yeh08}.
The synthesized beam 
size of the maps is shown in the bottom left corner of panel $d$.}
\label{Fig1}
\end{figure}

\begin{figure}[h]
\epsscale{1.0}
\includegraphics[width=\columnwidth]{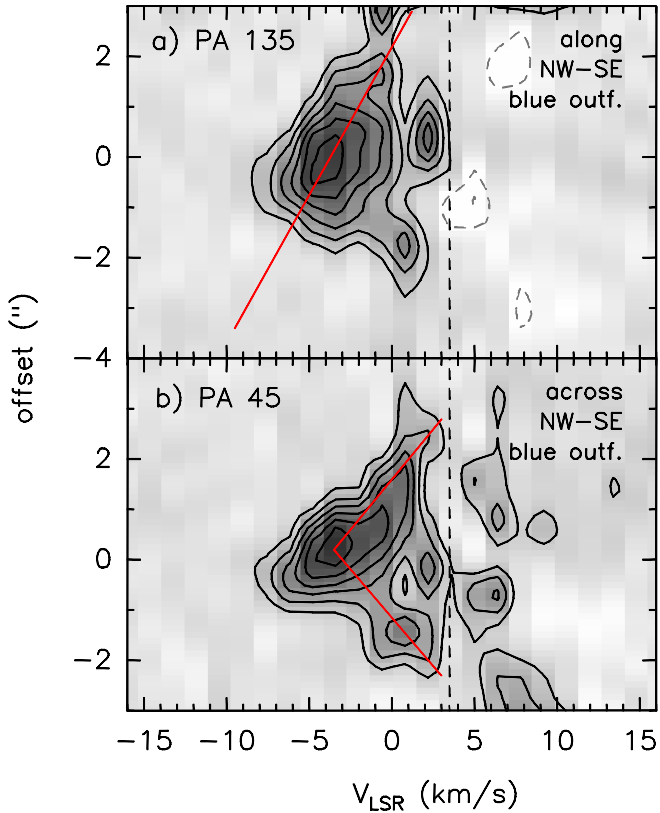}
\caption{Position--velocity plots of the CO 3--2 emission along (panel $a$) and 
across (panel $b$) the blueshifted lobe of the NW-SE compact outflow. The 
reference position ($0''$ offset) is 
RA(J2000.0)$=16^{\rm h}32^{\rm m}22\fs65$
and  DEC(J2000.0)$= -24\degr28\arcmin 33\farcs5$.  \irasA, the driving source 
of this outflow is located approximately at the top of
panel $a$ (offset position of $+3''$). The vertical dashed line indicates the 
systemic velocity of \irasA. The red lines are shown to better indicate the velocity 
gradient of the blueshifted CO emission.
}
\label{Fig2}
\end{figure}

\begin{figure}[h]
\epsscale{1.0}
\includegraphics[width=\columnwidth]{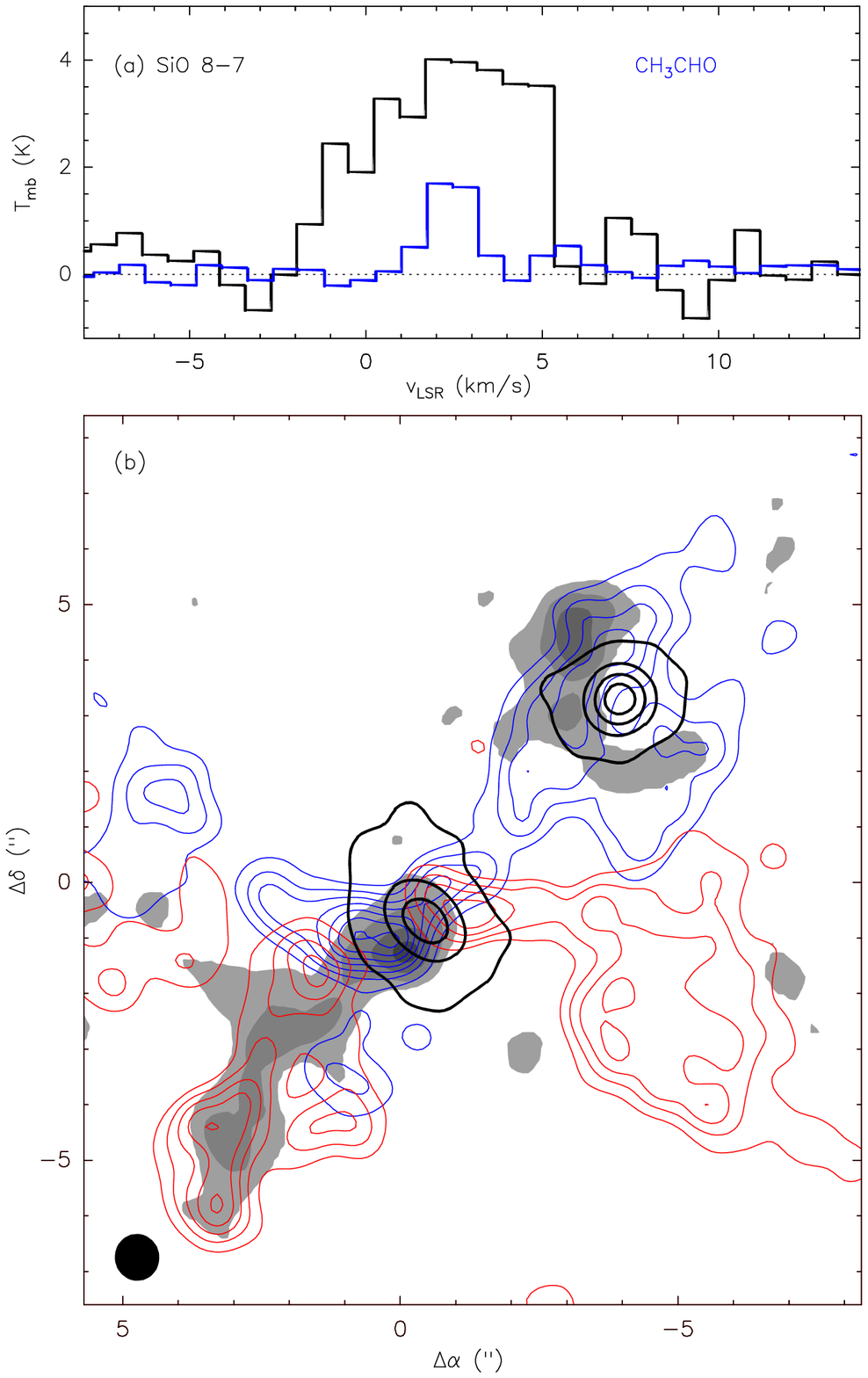}
\caption{
Panel $a$: Spectra of the SiO 8-7 (black spectrum) and CH$_3$CHO (blue 
spectrum) lines, averaged over an area of 9 arcsec$^2$ around \irasB. 
The CH$_3$CHO spectrum was obtained by  averaging the 
18$_{0,18}$-17$_{0,17}$ A and E, and  17$_{2,15}$-16$_{2,14}$ A and E lines
(these four lines are located  in the 334.9-335.4 GHz range and have similar
excitation temperatures and line strengths).
Panel $b$: Zoom in of the panel $d$ from Fig.~\ref{Fig1}
(contours of the CO and dust emission are the same as those from Fig.~\ref{Fig1}), 
overlapped of the integrated SiO 8-7 emission 
(grey image).
}
\label{Fig3}
\end{figure}

\begin{figure}[h]
\epsscale{1.0}
\includegraphics[width= \columnwidth]{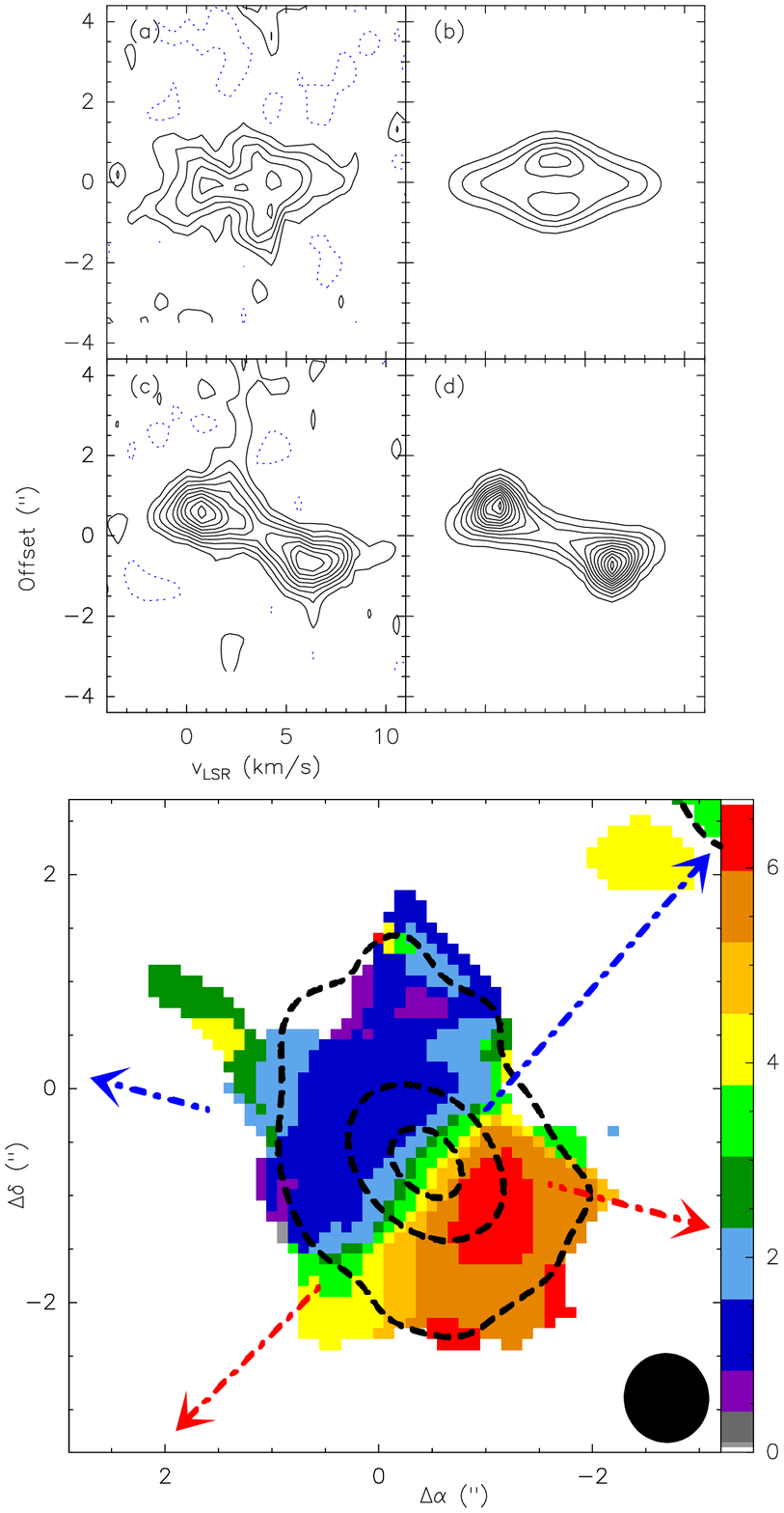}
\caption{Bottom panel: color image of the first order moment (velocity field) of
the \cts\ 7--6 line toward \irasA. The color scale (in \kms) is shown in the
right side of the panel. The black thick dashed contours show the 
878~$\mu$m continuum emission. The thick dashed blue and red arrows 
show the direction of outflows.  
Top  panels:
velocity--position plots of the \cts\ 7-6 lines taken along the minor (panel $a$) 
and major (panel $c$) axes. Panels $b$ and $d$ show the modeled data.
}
\label{Fig4}
\end{figure}

\section{Results}\label{res}

Figure~\ref{Fig1} shows the channel maps of the high velocity component of 
the CO 3--2 emission. This figure shows that the high velocity gas exhibits  a
quadrupolar morphology well centered in \irasA, apparently forming 
two bipolar molecular outflows, an extended E-W outflow and a compact
NW-SE outflow (the direction of these two outflows are delineated in 
Fig.~\ref{Fig1}$b$).  The E-W outflow consists of blueshifted emission in the 
eastern lobe (with some redshifted gas at low flow velocities) and redshifted emission 
in the western lobe.  This bipolar outflow has been previously well studied 
and extends to distances of 0.1~pc from its powering source \citep{Mizuno90, 
Hirano01, Yeh08}, far beyond the SMA field of view.  

The NW-SE bipolar outflow extends only $\simeq 8''$ (0.005~pc, 1000~AU), so it is
very compact compared with the other two bipolar outflows detected in the region
\citep{Mizuno90}.  This compact NW-SE bipolar outflow was already reported by 
\citet{Rao09}, but our new SMA observations now reveal it more clearly (in particular 
from the SiO data; see below).  At low outflow velocities the blueshifted lobe appears 
to split in two parts just before reaching \irasB\ (Fig.~\ref{Fig1}$c$ and 
\ref{Fig1}$d$).  At higher velocities (Fig.~\ref{Fig1}$a$ and Fig.~\ref{Fig1}$b$) 
the blueshifted gas ends 
at the position of \irasB.  Another characteristic of the outflow is an 
apparent acceleration of the gas, with the higher velocities appearing farther 
from the driving source. This is more evident in the position-velocity
cut along the blueshifted lobe (Fig.~\ref{Fig2}$a$): the terminal 
outflow velocity increases with the distance from the source. A cut across the 
blueshifted lobe (Fig.~\ref{Fig2}$b$) shows that the highest outflow 
velocities are spatially more compact than the outflow component at velocities 
closer to the systemic velocity of the \irasA core, suggesting that the outflow is 
more collimated at higher outflow velocities. 

The SiO 8-7 emission (Fig.~\ref{Fig3}) appears to arise mainly from two 
patches of emission. One of them extends from \irasA, in a jet-like structure, 
toward the south-east following well the redshifted lobe of the compact NW-SE 
outflow seen in CO 3-2.  Most of the emission in the jet-like structure appears 
close to the cloud velocity but slightly redshifted 
(\vlsr $\simeq 3.0$--$12.0$~\kms).  On the other hand, the northern patch of SiO 
emission shows a clear partial ring surrounding \irasB,  facing \irasAÊ
(Fig.~\ref{Fig3}), and appears near the position where the blue lobe of the CO 
compact NW-SE outflow diverges spatially (Fig.~\ref{Fig1}). 

The \cts\ 7-6 mainly traces the circumstellar gas around \irasA\ at scales of  few
hundreds AU and matches the 878~$\mu$m dust emission very well
(Fig.~\ref{Fig4}). Both the circumstellar molecular gas and the dust structures are  
elongated along the NE-SW direction ($PA \simeq 41\arcdeg$), with position 
velocity cuts of the CS emission along and across the major axis suggesting 
that the gas is rotating with a Keplerian-like pattern.  Below, we describe the 
procedure to fit the \cts\ 7--6 emission toward \irasA with a rotating geometrically 
thin disk. 

\subsection{Thin Disk Model for \irasA}

We considered a geometrically thin disk, with an inner ($r_{i}$) and outer
($r_{o}$) radius. The angle between the disk axis  and the plane of the sky is 
$i$ ($i=0^\circ$ for an edge-on disk).  We consider a rotation velocity 
given by a power law of the radius, $v_r (r/r_0)^{q_r}$, where $r_0$ is an 
arbitrary reference radius and $v_r$ is the rotation velocity at the reference 
radius.

We computed, for each point of a regular grid in the plane of the sky, the
projection of the rotation velocity of the corresponding point of the
disk along the line of sight $v_z$. A Gaussian line profile of width $\Delta v$
and centered on $v_z$ was added to the channels associated with the grid point.
Finally, each channel map was convolved spatially with a Gaussian beam of 
width $\Delta s$. However, the intensity scale of the channel maps is arbitrary. 
A scaling factor, the same for all channel maps, was obtained by minimizing the 
sum of the squared differences between the data channel maps  and the synthetic 
channel maps.
The model depends on a total of 10 parameters,  namely
the beamwidth, $\Delta s$;
the linewidth, $\Delta v$;
the disk center,  $(x_0, y_0)$; 
the disk systemic velocity, $v_0$; 
the disk inner and outer radii, $r_i$ and $r_o$; 
the disk rotation velocity at the reference radius, $v_r$;  
the radial dependence power-law index of the rotation 
velocity, $q_r$; and 
the disk inclination angle. $i$. 
Some of the parameters are known beforehand, such as $\Delta s$ and $\Delta v$.
Some other can be guessed based on physical grounds (i.e.\ $q_r=-0.5$).   These
considerations leave 7 free parameters, the first 3 are geometrical ($x_0$, $y_0$,
$v_0$), and the last 4 have with physical interest ($r_i$, $r_o$, $v_r$  and
$i$), which can be estimated through model fitting to the data.

The fitting procedure was the sampling of the seven-dimensional parameter space, 
using the same procedure as that described in \citet{Estalella12} and \citet{Sanchez13}.  
The parameter space was searched for the minimum value of the rms fit residual.  
Once a minimum of the rms fit residual was found, the uncertainty in the parameters 
fitted was found as the increment of each of the parameters of the fit necessary to 
increase the rms fit residual by a factor of
$[1+\Delta(m, \alpha)/(n-m)]^{1/2}$, where
$n$ is the number of data points fitted, 
$m$ is the number of parameters fitted, and 
$\Delta(m, \alpha)$ is the value of $\chi^2$ for $m$ degrees of freedom  (the number 
of free parameters) and $\alpha$ is the significance level ($0<\alpha<1$). For $m=7$, 
and for a significance level of 0.68 (equivalent to 1 $\sigma$ for a Gaussian error 
distribution), the increment in the rms fit residual is given by $\Delta(7,0.68)=8.17$
\citep{Sanchez13}.

\begin{table}
\caption{Parameters of the Best Fit Model
\label{tparams}}
\begin{tabular}{lcc}
\hline\hline
Parameter & Units & Value \\
\hline
{\sc Fixed:} \\
Beamwidth $\Delta s$		& (arcsec)		& $ 0.82$ \\
Linewidth $\Delta v$			& (km s$^{-1}$)	& $ 1.20$ \\
Rotation power-law index $q_r$&           	& $-0.50$ \\
{\sc Fitted:}\\
Disk center $x_0$			& (arcsec)		& $-0.12\pm0.02$ \\
Disk center $y_0$			& (arcsec)		& $ 0.20\pm0.01$ \\
Disk systemic velocity $v_0$	& (km s$^{-1}$)	& $ 3.49\pm0.09$ \\
Disk inner radius $r_i$		& (AU)			& $ 1\pm1$ \\
Disk outer radius $r_o$		& (AU)			& $ 140\pm2$ \\
Rotation velocity $v_r$~$^a$	& (km s$^{-1}$)	& $-6.5\pm0.2$ \\
Disk inclination $i$			& (deg)		& $44.2\pm0.9$ \\
\hline
\end{tabular}

$^a$ For a reference radius $r_0=0\farcs4$ (48 AU).
%\tablenotetext{a}{For a reference radius $r_0=0\farcs4$ (48 AU).}
\end{table}

The model was fitted to the \cts\ 7--6 emission associated to \irasA.  The 
rotation axis in the plane of the sky (derived from the CS velocity gradient) 
was found to be at a position angle of $-40\fdg8$.  The best fit values and their 
errors are shown in Table \ref{tparams}.  Figure~\ref{Fig4} shows the comparison 
of the synthetic position-velocity cuts for the best solution with those for the 
SMA \cts\ 7--6 data, with the best fit values matching the data very well.  The 
model used assumes that the intensity is proportional to the geometrical depth of 
the disk along the line of sight (i.e.\ optically thin emission, and constant 
density and temperature), which is reasonable for the \cts\ emission, except near 
the disk center. Thus, the non-zero value obtained  from the fit for the disk 
inner radius can be only interpreted in the sense that the contribution of the 
emission near the disk center is low.  The simple kinematical model used here 
cannot discard the idea that the inner radius is actually larger, as suggested by the 
binarity of the central source, with a semi-major axis of $0\farcs35$ 
\citep{Loinard07}.

The mass of the protostar in source A can be estimated from the values derived 
from the disk fitting. Assuming Keplerian rotation ($M = r \, v^2/G$), the mass
is $M_{\rm \irasA} = 2.3\pm0.1$~\msun. This mass is similar to the mass derived from 
the relative motions of the A1 and A2 objects \citep[][these two objects are 
embedded in the \cts\ structure]{Pech10}.

\section{Discussion and conclusions}

The first single-dish observations in I16293 showed two bipolar molecular outflows, 
one extended along the NE (redshifted)-SW (blueshifted) direction, and another along
the E (blueshifted)-W (redshifted) direction \citep{Walker88, Mizuno90}. The NE-SW 
outflow has not been detected at smaller scales through interferometric observations,
suggesting that it might be a fossil outflow \citep{Yeh08, Rao09}. The
E-W outflow has been well detected and studied with the SMA at arcsecond angular
resolution \citep{Yeh08, Rao09}. 
The CO emission associated with the western lobe at scales of $\sim 3000$~AU appears 
to follow a parabolic cavity with an inclination angle of $30\arcdeg$ with respect to 
the plane of the sky \citep[see the green dashed line in Figure~\ref{Fig1};][]{Yeh08}.  
However, our maps show that near the protostars the CO outflow 
emission does not follow  this parabolic cavity. 
At the position of \irasA\ the CO 3--2 outflow is very bright and extends roughly in
the direction of the E-W outflow.  At the highest velocity channel 
(Fig.~\ref{Fig1}$a$) the CO emission around \irasA\ is compact and the blue 
and redshifted peaks form a position angle of $\simeq100\arcdeg$, so they are 
possibly also associated with the E-W outflow.

The relatively high velocity CO 3--2 emission as traced by the SMA delineates 
a well defined bipolar compact outflow of only 0.005~pc in the NW-SE direction, 
well centered around \irasA, as was already suggested by \citet{Rao09}.  However, 
this differs from what has recently been reported from ALMA CO 6--5 observations 
\citep{Loinard13, Kristensen13}. There are some observational features that support 
our statement.  First, the NW-SE outflow appears to be parallel to the rotation axis 
of the circumstellar disk-like structure, traced by the \cts, around \irasA 
(see Fig.~\ref{Fig4}).  Second, the overall kinematical and morphological features of 
the NW blue lobe appear to be very consistent with those of the prototypical 
molecular outflows associated with class 0 protostars \citep[e.~g.,][]{Arce06, Palau06} 
if it is powered by \irasA: it has a conical-like structure starting in this source;
the CO 3-2 channel maps show an apparent Hubble-like velocity structure (higher 
velocities arise farther from the powering protostar: see also the position-velocity 
cut along the blue lobe in Fig.~\ref{Fig2}); a cut across the blue 
lobe shows (Fig.~\ref{Fig2}) that the highest CO 3-2 velocities occur along the outflow 
axis (\ie\ the highest velocities are more collimated than the lowest).  Third, the SiO 
emission is found only along the SE-NW direction. The SiO is a molecule that traces 
shocks strong enough to produce dust sputtering, 
releasing silicates from the dust mantles \citep[e.~g.,][]{Anderl13}. The morphology 
of the SiO emission  associated with \irasB\ suggests that the SiO arises from 
the external shells of the circumstellar material of this component. Its location, facing 
component \irasA\ and overlapping with the NW blueshifted lobe, suggests that
the SiO traces the region where the NW-SE outflow (powered by \irasA) is impacting on 
the circumstellar gas around \irasB.  In fact, the SiO spectrum of the
emission around \irasB is much broader than the acetaldehyde 
(CH$_3$CHO) spectrum (see Fig~\ref{Fig2}$b$). Acetaldehyde is a
hot core tracer and is likely tracing the quiescent (apparently  
unperturbed) circumstellar gas in \irasB.

The terminal velocity of this compact outflow is small, $\simeq 13$~\kms. 
This outflow is in projection perpendicular to the major axis of the disk-like 
structure associated with \irasA, so we can fairly assume that this configuration holds 
in 3-D. Thus correcting for the outflow inclination (44\arcdeg, see Table~\ref{tparams}), 
we find a dynamical timescale of $\sim 400$~yr.  This is much smaller 
than the kinematic timescale of the extended E-W outflow \citep[5000~yr,][]{Mizuno90}.  
We also estimated the outflow parameters of the compact NW-SE outflow following 
\citet{Palau07}, and assuming the same inclination given above, optically thin 
emission, and an excitation temperature of $\sim18$~K (derived from 
the line peak of the CO 3--2 spectrum).  We obtained a total mass of 
$\sim 2\times10^{-4}$~\msun, a mass outflow rate of $\sim 5\times10^{-7}$~\Mdot, and 
a momentum rate of $\sim 6\times10^{-6}$~\Pdot. The momentum 
rate and the bolometric luminosity of \irasA (somewhat smaller than $\simeq25$~\lsun, 
which is the total luminosity in the region) match the correlation between bolometric 
luminosity and outflow momentum rate found in the literature \citep{Beltran08, Takahashi12}.
Yet, the small outflow mass and dynamic time scales suggest that this is a very young
molecular outflow, possibly being powered by the youngest protostar in the region. 
Finally, we note that the water masers detected in the region appear to be redshifted 
and located only $0\farcs1$ (12 AU) South and South-East ($PA=157$-$182\arcdeg$) 
from source A1 \citep[][]{Pech10, Alves12}. This suggests that A1 could be the 
powering source of the compact outflow (its SE lobe is also redshifted).

In summary, the SMA data presented suggest a scenario where \irasA\ consists of 
at least two protostars, one driving the E-W outflow, and another driving a 
more compact and chemically rich NW-SE outflow (possibly A1). Both protostars 
in \irasA\ are embedded in a circumbinary disk traced by the 878~$\mu$m continuum 
and the C$^{34}$S 7--6 emission, which is elongated perpendicular to the NW-SE outflow 
and presents a velocity gradient also perpendicular to this outflow. This situation is similar 
to what is found in the intermediate-mass protostellar system IRAS\,22198+6336, where 
a binary is driving two perpendicular outflows, and is embedded in a disk rotating 
perpendicularly to the most chemically rich outflow  \citep{Sanchez10, Palau11}.  Finally, 
the NW-SE outflow driven by one of the protostars in \irasA seems to be impacting on 
\irasB, as revealed by the SiO 8--7 emission showing an arc-morphology at the 
border of \irasB\ facing \irasA. The physical and chemical effects of such an interaction 
on the dynamics and evolution of \irasB\ remain to be studied.

\acknowledgments The SMA, a joint project between the Smithsonian
Astrophysical Observatory  and the Academia Sinica  Institute of Astronomy and
Astrophysics, is funded  by the Smithsonian Institution and the Academia
Sinica. JMG also thanks the  SMA staff at Hilo for their support. RE, JMG, AP, 
and JMT are  supported by the Spanish MINECO AYA2011-30228-C03 and Catalan
AGAUR 2009SGR1172  grants. The ICC (UB) is a CSIC-Associated Unit through 
the ICE (CSIC).


\begin{thebibliography}{}

\bibitem[Alves et al.(2012)]{Alves12}
 Alves, F.~O., Vlemmings, W.~H.~T., Girart, J.~M., \& Torrelles, J.~M.\ 2012,
 \aap, 542, A14 
 
\bibitem[Anderl et al.(2013)]{Anderl13}
  Anderl, S., Guillet, V., Pineau des For{\^e}ts, G., \& Flower, D.~R.\ 2013, \aap, 556, A69 

\bibitem[Arce \& Sargent(2006)]{Arce06}
 Arce, H.~G., \& Sargent, A.~I.\ 2006, \apj, 646, 1070 

\bibitem[Beltr{\'a}n et al.(2008)]{Beltran08}
 Beltr{\'a}n, M.~T., Estalella, R., Girart, J.~M., Ho, P.~T.~P., \& Anglada, G.\ 
 2008, \aap, 481, 93 

\bibitem[Chandler et al.(2005)]{Chandler05}
 Chandler, C.~J., Brogan, C.~L., Shirley, Y.~L., \& Loinard, L.\ 2005, \apj,
 632, 371 

\bibitem[Estalella et al.(1991)]{Estalella91}
 Estalella, R., Anglada, G., Rodriguez, L.~F., \& Garay, G.\ 1991, \apj, 371,
 626 
 
\bibitem[Estalella et al.(2012)]{Estalella12}
 Estalella, R., L\'opez, R, Anglada, G., G\'omez, G., Riera, A., \&
 Carrasco-Gonz\'alez, C.  2012, \aj, 144, 61
   
 \bibitem[Hirano et al.(2001)]{Hirano01}
  Hirano, N., Mikami, H., Umemoto, T., Yamamoto, S., \& Taniguchi, Y.\ 2001,
 \apj, 547, 899 

\bibitem[Imai et al.(2007)]{Imai07}
 Imai, H., Nakashima, K., Bushimata, T., et al.\ 2007, \pasj, 59, 1107 

\bibitem[Knude \& Hog(1998)]{Knude98}
 Knude, J.,  \& Hog, E. 1998, A\&A, 338, 897

\bibitem[Kristensen et al.(2013)]{Kristensen13}
 Kristensen,ÊL.~E., Klaassen,ÊP.~D., Mottram,ÊJ.~C., Schmalzl,ÊM., \&
 Hogerheijde,ÊM.~R.\ 2013, A\&A, 549, L6

 \bibitem[Loinard et al.(2007)]{Loinard07}
  Loinard, L., Chandler, C.~J., Rodr{\'{\i}}guez, L.~F., et al.\ 2007, \apjl, 670, 1353

\bibitem[Loinard et al.(2008)]{Loinard08}
 Loinard, L., Torres, R.~M., Mioduszewski, A.~J., \& Rodr{\'{\i}}guez, L.~F.\ 
 2008, \apjl, 675, L29 

\bibitem[Loinard et al.(2013)]{Loinard13}
 Loinard, L., Zapata, L.~A., Rodr{\'{\i}}guez, L.~F., et al.\ 2013, \mnras, 430,
 L10 

\bibitem[Lombardi et al.(2008)]{Lombardi08}
 Lombardi, M., Lada, C.~J., \& Alves, J.\ 2008, \aap, 480, 785 

\bibitem[Mizuno et al.(1990)]{Mizuno90}
 Mizuno, A., Fukui, Y., Iwata, T., Nozawa, S., \& Takano, T.\ 1990, \apj, 356,
184 

\bibitem[Palau et al.(2007)]{Palau07}
 Palau, A., Estalella, R., Ho, P.~T.~P., Beuther, H., \& Beltr{\'a}n, M.~T.\ 2007, 
 \aap, 474, 911 

\bibitem[Palau et al.(2011)]{Palau11}
 Palau, A., Fuente, A., Girart, J.~M., et al.\ 2011, \apjl, 743, L32 

\bibitem[Palau et al.(2006)]{Palau06}
 Palau, A., Ho, P.~T.~P., Zhang, Q., et al.\ 2006, \apjl, 636, L137 

\bibitem[Pech et al.(2010)]{Pech10}
 Pech, G., Loinard, L., Chandler, C.~J., et al.\ 2010, \apj, 712, 1403 
 
 \bibitem[Pineda et al.(2012)]{Pineda12}
  Pineda, J.~E., Maury, A.~J., Fuller, G.~A., et al.\ 2012, \aap, 544, L7 

\bibitem[Rao et al.(2009)]{Rao09}
 Rao, R., Girart, J.~M., Marrone, D.~P., Lai, S.-P., \& Schnee, S.\ 2009, \apj,
 707, 921 

\bibitem[Rao et al.(2013)]{Rao13}
  Rao, R., Girart, J.~M., Lai, S.-P., \& Marrone, D.\ 2013, \apjl, in press, arXiv:1311.6225

\bibitem[Rodr{\'{\i}}guez et al.(2005)]{Rodriguez05} 
 Rodr{\'{\i}}guez, L.~F., Loinard, L., D'Alessio, P., Wilner, D.~J., \& Ho,
 P.~T.~P.\ 2005,  \apjl,  621, L133 

\bibitem[S{\'a}nchez-Monge et al.(2010)]{Sanchez10}
 S{\'a}nchez-Monge, {\'A}., Palau, A., Estalella, R., et al.\ 2010, \apjl, 721, L107 

\bibitem[S\'anchez-Monge et al.(2013)]{Sanchez13}
 S\'anchez-Monge, \'A., Palau, A., Fontani, F., et al.\ 2013, \mnras, 432, 3288 

\bibitem[Takahashi \& Ho(2012)]{Takahashi12}
 Takahashi, S., \& Ho, P.~T.~P.\ 2012, \apjl, 745, L10 

\bibitem[Walker et al.(1988)]{Walker88}
 Walker, C.~K., Lada, C.~J., Young, E.~T., \& Margulis, M.\ 1988, \apj, 332,
 335 

\bibitem[Wilking \& Claussen(1987)]{Wilking87}
 Wilking, B.~ A., \& Claussen, M.~J.\ 1987, \apj, 320, L133

\bibitem[Yeh et al.(2008)]{Yeh08}
 Yeh, S.~C.~C., Hirano, N., Bourke, T.~L., et al.\ 2008, \apj, 675, 454

\bibitem[Wootten(1989)]{Wootten89}
 Wootten, A.\ 1989, \apj, 337, 858 

\bibitem[Wright  \& Sault(1993)]{Wright93}
 Wright, M.~C.~H., \& Sault, R.~J.\ 1993, \apj, 402, 546 

\bibitem[Zapata et al.(2013)]{Zapata13}
 Zapata, L.~A., Loinard, L., Rodr{\'{\i}}guez, L.~F., et al.\ 2013, \apjl, 764,
L14 

\end{thebibliography}
\end{document}